\documentclass{article}
\usepackage{mdframed}
\usepackage[most]{tcolorbox}
\usepackage{todonotes}
\usepackage{PRIMEarxiv}
\usepackage[utf8]{inputenc} 
\usepackage[T1]{fontenc}    
\usepackage{hyperref}  
\usepackage{amssymb,amsmath,amsthm}
\usepackage{xspace}
\usepackage{mathtools}
\usepackage{graphicx,xcolor}
\usepackage{url}            
\usepackage{booktabs}       
\usepackage{amsfonts}       
\usepackage{nicefrac}       
\usepackage{microtype}      
\usepackage{lipsum}
\usepackage{fancyhdr}       
\usepackage{graphicx} 
\usepackage[utf8]{inputenc}
\graphicspath{{media/}}     

\tcbuselibrary{skins}
\tcbuselibrary{breakable}

\newtcolorbox{coloredframe}[3][]{
    empty,
    breakable=true,
    sharp corners=all,
    top=4mm, left=4mm,
    borderline west={1.5pt}{0pt}{#3}, borderline north={1.5pt}{0pt}{#3},
    attach boxed title to top left={yshift=-1.75ex,xshift=6ex},
    coltitle=black,
    colback=white, colbacktitle=white,
    fonttitle=\bfseries,
    boxed title style={boxrule=0pt,colframe=white},
    title=#2,
    #1
}
\usepackage{lineno}

\newtheorem{theorem}{Theorem}[section]
\newtheorem{corollary}{Corollary}[theorem]
\newtheorem{lemma}[theorem]{Lemma}
\newtheorem{definition}{Definition}[section]
\title{On Approximating the Dynamic and Discrete Network Flow Problem
}

\author{
Bubai Manna\thanks{Department of Mathematics, Indian Institute of Technology Kharagpur, India. \texttt{bubaimanna11@gmail.com}} \And Bodhayan Roy\thanks{Department of Mathematics, Indian Institute of Technology Kharagpur, India. \texttt{bodhayan.roy@gmail.com}} \And Vorapong Suppakitpaisarn \thanks{Graduate School of Information Science and Technology, The University of Tokyo, Japan. \texttt{vorapong@is.s.u-tokyo.ac.jp}} 
}

\begin{document}
\maketitle

\begin{abstract}
We examine the dynamic network flow problem under the assumption that the flow consists of discrete units. The dynamic network flow problem is commonly addressed in the context of developing evacuation plans, where the flow is typically treated as a continuous quantity. However, real-world scenarios often involve moving groups, such as families, as single units. We demonstrate that solving the dynamic flow problem with this consideration is APX-hard. Conversely, we present a PTAS for instances where the base graph is a path with a constant number of nodes. We introduce a `ready time' constraint to the minsum bin packing problem, meaning certain items cannot be placed in specific bins, develop a PTAS for this modified problem, and apply our algorithms to the discrete and dynamic flow problem.
\vspace{-0.1in}
\end{abstract}


\section{Introduction}
\label{sec:typesetting-summary}

Flow problems on dynamic graphs \cite{ford1958constructing} are considered by many researchers (e.g. \cite{higashikawa2022almost, klinz2004minimum, HIGASHIKAWA202187}) because of many reasons. One of the reasons is their relevance in finding evacuation routes during emergencies such as earthquakes or fires \cite{higashikawa2019survey}. In those applications, we aim to move persons in the ways that they arrive at aiding facilities as soon as possible. 
The most common objective function for those problems is minmax, which aims to minimize the time until all persons arrive at facilities. In this work, however, we consider another common objective function called minsum, which aims to minimize the summation of time that each individual needs for their trips.

In 1958, the paper \cite{ford1958constructing} solved the problem \emph{Maximum Flow Over Time}, while their `minmax' variation requested for \emph{Quickest Transshipment}, which was even studied lately in \cite{skutella2023introduction}. It has been shown that both objective functions of flow problems can be solved using time-expanded networks \cite{ford1958constructing,fu2016clustering}. However, these temporal graphs can be exponentially large in relation to the input size, making the algorithm pseudo-polynomial. 
For minmax problems, polynomial-time algorithms have been developed for paths \cite{bhattacharya2017improved} and trees \cite{treeminmax}. There are also FPTAS for general graphs when the number of facilities is constant \cite{belmonte2015polynomial}. In contrast, minsum problems have only been shown to have polynomial-time algorithms for path graphs \cite{benkoczi2020minsum}. 

All known algorithms assume that individuals are distinct and identical, meaning that we can move any number of people over a particular edge as long as the total number does not exceed the edge's capacity. However, this may not always be possible in practice. For example, some groups of people, such as families, must be moved together, and some groups may require emergency aid and should therefore be given higher priority. These considerations must be taken into account when determining how to move people from one location to another.

\subsection{Our Contributions}

We define our problem called minsum problem for discrete and weighted set flow on a dynamic network (MS-DWSF) in subsection \ref{subsection1.2}. It is clear that the discrete and weighted version is harder than the original problem. Indeed, it is easy to show that it is NP-hard by a reduction to the partition problem\footnote{The reduction has been given in our paper under submission.}. 

In Section \ref{section2}, the APX-hardness of MS-DWSF is established. Furthermore, Section \ref{PTAS} introduces a Polynomial Time Approximation Scheme (PTAS) specifically for instances where the underlying graph is a path graph featuring a single facility, a constant number of nodes, and uniform edge capacities. It's important to highlight that this assumption of uniformity and simplicity in graph structure is similarly adopted in several studies within the realm of dynamic network flows \cite{higashikawa2014minimax,kamiyama2006efficient, fujie2021minmax,higashikawa2015minimax, Skutella2009}.


We find that in the scenario where the graph forms a path with a single facility, the edge nearest to the facility experiences the highest congestion. Consequently, the challenge of dynamic flow becomes a matter of strategizing the distribution of discrete flows across this edge. This scenario aligns with the principles of the minsum bin packing problem \cite{epstein2007minimum,epstein2018min}, where the dispatch of items at a specific time $t$ can be likened to packing these items into a bin labeled $t$. Nonetheless, a direct application of bin packing algorithms is not feasible. This is because items originating from nodes distant from the facility require time to approach the nearest node, implying that these items cannot be dispatched prior to a specific time $\tau$. In essence, items cannot be ``packed'' in a bin numbered less than $\tau$. This observation has led us to explore the minsum bin packing problem with a consideration of ``ready time'' in our study.

In brief, for each item $i$, we assign a ``ready time'' $\tau_i$, indicating that the item cannot be placed in any bin numbered lower than $\tau_i$. We develop a Polynomial Time Approximation Scheme (PTAS) for this modified problem, which, in turn, provides a PTAS for the dynamic network flow problem described earlier. While we draw upon algorithmic concepts from prior research on the minsum bin packing problem, extending these ideas to our context is not straightforward. Consequently, this work introduces several new theoretical contributions.

\subsection{Problem Definitions}\label{subsection1.2}
In this section, we define our problem called minsum problem for discrete and weighted set flow on a dynamic network (MS-DWSF). 

Consider a graph with $n$ nodes, denoted by $V = \{1, \dots, n\}$ and $E \subseteq \{\{u,v\}, u,v \in V\}$. Each node $i$ has $m_i$ sets of persons to evacuate. These sets of persons are denoted by $G_{i,1}, \dots, G_{i,m_i}$. For group $G$, the size of $G$ is denoted by $S(G) \in \mathbb{Z}_{+}$ and the weight of $G$ is denoted by $w(G) \in \mathbb{Z}_{+}$. The capacity of edge $e$ is $C(e) \in \mathbb{Z}_{+}$. In addition, each edge $e$ has distance $d(e) \in \mathbb{Z}_{+}$, which is the time that persons need to move between two terminals of the edge. 


Let us denote the collection of groups that are at node $i$ at time $t$ by $\mathcal{S}_i^{(t)}$. We select from $\mathcal{S}_i^{(t)}$ which groups to be sent along the edge $\{i, j\} \in E$. We denote the collection of groups that we choose to send along $e = \{i,j\}$ by $D_{i,j}^{(t)}$. The summation of group sizes in $D_{i,j}^{(t)}$ must not be larger than $C$, i.e. $\sum\limits_{G \in D_{i,j}^{(t)}} S(G) \leq C(\{i,j\})$.

For $t = 0$, we have $S_i^{(0)} = \{G_{i,1}, \dots, G_{i,m_i}\}$ for all $i$.
Let denote $A_{j,i}^{(t)}$ be a collection of groups arriving at $i$ from node $j$ at time $t$. We have
$$A_{j,i}^{(t)} = \begin{cases}
  D_{j,i}^{(t - d(\{j, i\}))} & \{j,i\} \in E \text{ and } t \geq d(\{j,i\}) \\
  \emptyset & \text{ otherwise, and}
\end{cases}$$
Then, $S_i^{(t)} = S_i^{(t - 1)} \backslash \left( \bigcup\limits_{j:\{i,j\}\in E} D_{i,j}^{(t)} \right) \cup \left( \bigcup\limits_{j:\{i,j\}\in E} A_{j,i}^{(t)} \right)$.

Let the set of facilities be $F \subseteq V$.
The arrival time of $G$, denoted by $\alpha(G)$ is the smallest time of arrival that the group is at some node in $F$, i.e. $\min \left\{t: G \in \bigcup\limits_{i \in F} S_i^{(t)}\right\}$. In the MS-DWSF, we aim to minimize $\sum\limits_G w(G)\alpha(G)$.

\section{APX-hardness}\label{section2}
\begin{figure}[t]
\includegraphics[width=14cm]{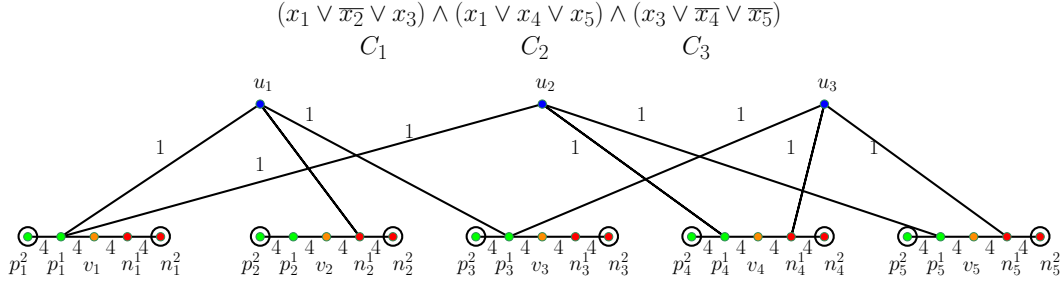}
\centering
\caption{The reduction graph for a given 4OCC-3SAT formula.}\label{reduction2}
\end{figure}

In this section we show that MS-DWSF is APX-hard when the underlying graph is bipartite.
We reduce from Max-4OCC-3SAT. This is a version of 3SAT where every variable occurs at most four times in the formula, and we need to satisfy as many clauses as possible. It is known to be APX-hard \cite{ECCC-TR03-022}. We start from a given 4OCC-3SAT formula $\theta$ on $n$ variables $x_1, x_2, \ldots, x_n$ and $m$ clauses $C_1, C_2, \ldots, C_m$. We construct an instance of MS-DWSF in the following steps (see Figure \ref{reduction2}):
\begin{enumerate}
    \item Initiate an empty graph $G(V,E)$.
    \item For every variable $x_i \in \theta$, add a vertex $v_i$ to $V$ (called a \emph{variable vertex}).
    \item For every clause $C_j \in \theta$, add a vertex $u_j$ to $V$
    (called a \emph{clause vertex}).
    \item For every vertex $v_i \in V$, add four vertices $p_i^1$, $p_i^2$, $n_i^1$ and $n_i^2$ to $V$, and place facilities on vertices $p_i^2$ and $n_i^2$.
    \item If $x_i \in C_j$, then add to $E$ the edge $(u_j, p_i^1)$ with capacity $1$ and distance $1$. Similarly, if  $\overline{x_i} \in C_j$, then add to $E$ the edge $(u_j, n_i^1)$ with capacity $1$ and distance $1$. 
    \item For all $i \in [1,n]$, add to $E$ the four edges $(x_i,p_i^1)$, $(x_i,n_i^1)$, $(p_i^1,p_i^2)$ and $(n_i^1,n_i^2)$, with capacity $4$ and distance $1$ each.
    \item On each $v_i$, place a group $g_i$ with size $4$ and weight $5$. On each $u_j$, place a group $h_j$ with size $1$ and weight $1$
\end{enumerate}
We have the following lemma.
\begin{lemma}
There is a constant $c$ such that MS-DWSF cannot be approximated better than a factor of $(1+c)$ in polynomial time.
\end{lemma}
\begin{proof} 
Consider any scheduling on the constructed graph. Each facility is at distance $2$ from the unit weighted groups of at most $4$ clause vertices, since we reduced from a 4OCC-3SAT formula. On the other hand, the groups at the variable vertices each carry a weight $5$. So,
to optimize the minsum, the groups from every variable vertex must reach the corresponding facilities at time $2$, contributing $10n$ to the minsum. Since these groups weigh $4$ units, this stops the unit weight groups from reaching the same facilities at time $2$. Thus it is clear that such a scheduling actually corresponds to an assignment to the underlying 3SAT formula; if the group at $v_i$ goes to the facility on $n^2_i$, then $x_i$ is assigned $1$, and $0$ otherwise. All the clause vertices whose groups can reach facilities at time $2$ correspond to satisfied clauses. 
Suppose that $k$  groups from the clause vertices reach facilities at time $2$. Then they contribute $2k$ to the minsum and the rest of the groups contribute $3m-3k$, making the minsum a total of $10n + 3m -k$. 
Since it is known that at least half of all clauses can be satisfied in a 3SAT formula, we have $k \geq  m/2$ for the optimum solution. Also, $3m \geq n$ in any 3SAT formula. Substituting, we get $10n + 3m -k \leq 65k$. 
So, any solution that is at most $(1+c)$ times the optimum minsum, is at most $65(1+c)k$. This corresponds to satisfying at most $65ck$ less clauses in the 3SAT formula. From \cite{ECCC-TR03-022}, we know that Max-4OCC-3SAT cannot be approximated better than by a factor of $1/1900$. Setting $c = \frac{1}{65\times 1900}$ we have the desired result.

 \end{proof}
\begin{corollary}
MS-DWSF is APX-hard even when the underlying graph is bounded degree, bipartite, with all edges having the same length.
\end{corollary}
\begin{proof}
Due to our construction, $G$ is bipartite, and all edges have the same length. Since the reduction preserves approximation, the claim follows.
\end{proof}

\section{PTAS}\label{PTAS}

In this section, we explore a specific instance of the MS-DWSF, focusing on a path graph $(V,E)$ with a fixed number of nodes, denoted as $V = \{1, \dots, k\}$ and $E = \{\{1,2\}, \dots, \{k-1,k\}\}$. We assume uniform capacity across all edges, represented by $C$. Furthermore, this scenario includes a single facility located at node $1$, i.e., $F = \{1\}$. Extending the algorithm designed for a facility at node $1$ to any other node is relatively straightforward. Assuming the facility is positioned at node $i$, the task of transporting items from nodes $\{1, \dots, i - 1\}$ to node $i$ and from nodes $\{i + 1, \dots, r\}$ to node $i$ becomes two separate challenges that can be independently addressed using the same algorithm.

It is evident that the most critical edge in our instance is the edge $\{1,2\}$. If we have a way to transport items in this edge, i.e. we know the set $D_{2,1}^{(t)}$ for all $t$, then, for $i > 1$, we can have $D_{i + 1,i}^{(t - \sum_{i' = 2}^{i} d(i' + 1, i'))} = D_{2,1}^{(t)} \cap (\bigcup\limits_{i' > i} S_{i'}^{(0)})$. We then can reduce the MS-DWSF to the bin packing problem, where we need to pack all items into the bins of size $C$, which are $D_{2,1}^{(1)}, \dots, D_{2,1}^{(T)}$ where $T$ indicates the maximum time bins are packed. From now, we will call the collection $D_{2,1}^{(t)}$ as bin $t$. 

Our problem then looks similar to the minsum bin packing problem. However, certain items, such as item $G$, may not be eligible for inclusion in some bin $D_{2,1}^{(t)}$. For instance, if $G$ originates from $S_3^{(0)}$, it cannot be assigned to $D_{2,1}^{(0)}$, signifying it is not feasible to transport $G$ across the edge $\{2,1\}$ at time $0$. This is due to the requisite time steps needed for $G$ to move from node $3$ to node $2$. Consequently, item $G$ cannot be allocated to the set $D_{2,1}^{(t)}$ for any $t < d(3,2)$, reflecting the inherent delay in moving items through the graph. 

In our approach, we adapt the minsum bin packing framework to include a `ready time' $\tau(G)$ for each item $G$, ensuring that the item cannot be placed into any bin numbered less than $\tau(G)$. Given that our model operates on a path graph with a fixed number of nodes, the set of possible $\tau(G)$ values is finite and of constant size, represented by $\{\tau_1, \dots, \tau_k\}$. We work under the assumption that these ready times are ordered such that $\tau_1 < \cdots < \tau_k$. Therefore, the problem and objective function for the polynomial-time approximation scheme are defined as follows.

\begin{tcolorbox}[breakable,bicolor,
  colback=cyan!5,colframe=cyan!5,boxrule=0pt,frame hidden]

\noindent\textbf{\underline{Bin-Packing and Objective Function:}}
\vspace{4pt}
One end of the dynamic edge refers to the position of items, while the other end refers to the \emph{facility location} as previously stated. Assume that we have sets of items $G_1,\cdots, G_k$ such that $\tau(G_i)=\tau_i$ and $\tau_1 < \cdots < \tau_k$. We also assume that we have a sufficient number of bins, and each bin may hold a maximum of $1$ in unit weight. It is important to note that the weight of any item cannot exceed $1$ unit; otherwise, it cannot be packed. The dynamic edge can only hold one bin per unit of time. That is, we can pack exactly one bin per unit of time. If the bin $B_i$ is packed at time $i$, then the cost to transport $B_i$ on the dynamic edge is $i\times \lvert B_i\rvert$, where $\lvert B_i\rvert$ is the cardinality of the bin $B_i$. When time $t<\tau_1$, there are no items ready for packing because $\tau_1 < \cdots < \tau_k$. Therefore, the bins $B_1$, $B_2$,\dots, $B_{\tau_1-1}$ are empty. When $t = \tau_1$, the items of $S_1$ (we assume $S_1=G_1$ initially) are only ready for packing until $t = \tau_2$. We pack one bin per unit of time; hence, we need $\tau_2-\tau_1$ bins before $t=\tau_2$. After packing the bins $B_{\tau_1}$, $B_{\tau_1+1}$,\dots, $B_{\tau_2-1}$, when we arrive at time $t=\tau_2$, we have some unpacked items of $S_1$ and the items of $G_2$ to be ready. Therefore, when $t=\tau_2$, the items of $S_2$ (we assume $S_2=G_2\cup S_1\setminus (\cup_{i=\tau_1}^{\tau_2-1}B_i)$) are ready to be packed. We pack an additional $\tau_3-\tau_2$ bins before $t=\tau_3$. We repeat the process until $t =\tau_k-1$. When we reach $t=\tau_k$, the items of $S_k$ (we assume $S_k=G_k\cup S_{k-1}\setminus (\cup_{i=\tau_{k-1}}^{\tau_k-1}B_i)$) are ready to be packed. Finally, pack all of the items of $S_k$ using a minimum number of bins to minimize $\sum_{j=1}^{t} j\times \lvert B_j\rvert$ where $B_t$ is the last bin packed to evacuate all the items to the facility location (that is, the other end of the dynamic edge). Therefore, our target is to minimize the cost $\sum_{j=1}^{t} j\times \lvert B_j\rvert$. A bin $B_j$ is called a \emph{feasible bin} if the total weight of the items packed in $B_j$ is less than or equal to $1$. The feasible bin is referred to as an \emph{optimal bin} if it appears in an optimal solution. A packing is called \emph{feasible packing} if all the bins of that packing are feasible. Let $n$ be the total number of items in $\cup_{i=1}^{k}G_i$.
\end{tcolorbox}

\subsection{Properties of Items, Their Weights and the Partition}\label{properties}

We select a fixed constant $\epsilon$ with some properties and weights indicated below.

\begin{enumerate}
\item $0 < \epsilon \leq \frac{1}{4}$ be a fixed constant such that $\frac{1}{\epsilon}$ is an integer.
\item An item is said to be \emph{small} if its weight does not exceed $\epsilon^2$; otherwise, it is classified as \emph{large}.
\item $\lvert G_i\rvert =n_i$ $\mbox{ for } i=1,\dots,k$. Therefore $n=\sum_{i=1}^{k}n_i$.
\item It is assumed that $G_i$ contains a set of $m_i$ large items, each of which is contained in the set $L_i=\{x_{i,n_i-m_{i}+1}, x_{i,n_i-m_i+2},\dots,x_{i,n_i}\}$ with $w_{i,n_i-m_i+1} \leq w_{i,n_i-m_i+2} \le \dots \leq w_{n_i}$. Hence, $m=m_1+m_2+ \dots +m_k$, where $m$ is the total number of large items in $I$.
    
\end{enumerate}
As a result, we have a set of items $S_i$ at time $t=\tau_i$ and $\tau_{i+1}-\tau_i$ ($i\neq k$) bins in which to place some items of $S_i$ in order to accommodate the greatest possible number of items. Additionally, when $i=k$, a minimum number of bins is required to store all the items of $S_k$. Consequently, in order to store items of each $S_i$, a polynomial-time approximation scheme is established below.

\begin{lemma}\label{bubai17}
Without loss of generality, we can assume that at least one $n_i\geq \frac{1}{\epsilon^3}$ $\mbox{ for } i=1,\dots,k$.
\end{lemma}
\begin{proof}
If the claim is invalid, then $n_i < \frac{1}{\epsilon^3}$ for all $i = 1, 2, \dots, n$. For an arbitrary bin $B_j$, $B_j$ must pack some items of one of the sets $S_1$, $S_2$, \dots, $S_k$. If each $n_i<\frac{1}{\epsilon^3}$, then size of each $S_i$ is bounded by $\frac{k}{\epsilon^3}$ $\mbox{ for } i=1,\dots,k$. Consequently, the cardinality of the power set of each $S_i$ must be bounded by $2^{\frac{k}{\epsilon^3}}$ and at least one set of that power set corresponds to the bin $B_j$. $\tau_k+n$ represents the utmost number of bins that can be accommodated in the optimal solution. Therefore, the optimal solution can be determined in the time $\mathcal{O}((n+\tau_k)2^{\frac{k}{\epsilon^3}})\leq \mathcal{O}((\frac{k}{\epsilon^3}+\tau_k)2^{\frac{k}{\epsilon^3}})$ where $k$ and $\tau_k$ are constants.
\end{proof}

\smallskip
\begin{tcolorbox}[breakable,bicolor,
  colback=cyan!5,colframe=cyan!5,boxrule=0pt,frame hidden]

\noindent \textbf{\underline{Partition:}}\label{partition}
\vspace{4pt}
We divide $L_i$ into $\frac{1}{\epsilon^3}$ subsets denoted as $L_{i,1}$, $L_{i,2}$, \dots, $L_{i,\frac{1}{\epsilon^3}}$, based on two properties listed below.  
\begin{enumerate}
    \item  $\lvert L_{i,1} \rvert \geq \lvert L_{i,2} \rvert \geq \dots \geq \lvert L_{i,\frac{1}{\epsilon^3}} \rvert \geq \lvert L_{i,1} \rvert -1$ (this indicates that the cardinality of each subset is nearly identical to that of $\epsilon^3\lvert L_i \rvert$).
    
    \item If $x_{i,a}\in L_{i,j}$ and $x_{i,b}\in L_{i,s}$ such that $j < s$, then $a > b$ and $w_{i,a} \geq w_{i,b}$ (this means that the largest $\lvert L_{i,1}\rvert$ items of $L_i$ are assigned to $L_{i,1}$, followed by the largest $\lvert L_{i,2}\rvert$ items of remaining items of $L_i$ assigned to $L_{i,2}$ and so on).
\end{enumerate}
\end{tcolorbox}

The first property of the partition determines the exact cardinality of the $L_{i,j}$ which is either $\lfloor \epsilon^3\lvert L_i \rvert \rfloor$ or $\lceil \epsilon^3\lvert L_i \rvert \rceil$ $\mbox{ for } j=1,\dots, \frac{1}{\epsilon^3}$ and the partition is well-defined by the second property.

\begin{lemma}{\label{bubailemma2}} The following inequalities hold:
\begin{enumerate}
    \item $\lvert G_i \setminus L_{i,1} \rvert \geq n_i(1-2\epsilon^3)$
    \item  $\lvert L \rvert \leq 2m\epsilon^3$ and $\lvert I \setminus L \rvert \geq n(1-2\epsilon^3)$ where $L=\cup_{i=1}^{k}L_{i,1}$.
\end{enumerate}
   
\end{lemma}
\begin{proof}
$\lvert L_{i,1}\rvert$ has a weight of either $\lfloor \epsilon^3\lvert L_i \rvert \rfloor$ or $\lceil \epsilon^3\lvert L_i \rvert \rceil$. Consequently, $\lvert L_{i,1}\rvert \leq 2m_i\epsilon^3$ since $\lvert L_{i}\rvert =m_i$. We obtain $\lvert L\rvert \leq 2m\epsilon^3$ by taking the summation on both sides. Now, $\lvert I\setminus L\rvert \geq n-2m\epsilon^3 \geq n(1-2\epsilon^3)$ since $n\geq m$. Again, $\lvert G_i\setminus L_{i,1}\rvert \geq n_i-2m_i\epsilon^3 \geq n_i(1-2\epsilon^3)$ since $n_i\geq m_i$.

\end{proof}

\begin{tcolorbox}[breakable,bicolor,
  colback=cyan!5,colframe=cyan!5,boxrule=0pt,frame hidden]

\noindent \textbf{\underline{Rounded-up Instance:}}\label{roundedupsize}
\vspace{4pt}
For all $j\geq 2$, we assign the largest item in $L_{i,j}$ as the \emph{rounded up weight} of each item in $L_{i,j}$. This means that for all $j\geq 2$ and $x_{i,r}\in L_{i,j}$, we allow $\sigma_{i,j}=\max_{x_{i,r}\in L_{i,j}}w_{i,r}$ be the rounded-up weight of all the items of $L_{i,j}$. Despite this, the weights of the items in $L_{i,1}$ remain unchanged. There are no changes to the weight of the items in $G_i\setminus L_i$ (that is, the small items). We say that an item $x_i\in G_i\setminus L_i$ has weight $s_i$. The two conditions of the partition \ref{partition} must undoubtedly remain valid even after modifying the item weights in each $L_{i,j}$.

\smallskip
\end{tcolorbox}
\begin{tcolorbox}[breakable,bicolor,
  colback=cyan!5,colframe=cyan!5,boxrule=0pt,frame hidden]

\noindent \textbf{\underline{Packing Instance of OPT:}}\label{packinginstance}
\vspace{4pt}
Consider $B_1^{OPT}$, $B_2^{OPT}$, \dots, $B_{q_t}^{OPT}$ to be the optimal bins packed at times $t=0$, $t=1$, \dots, $t=q_t$, respectively, and OPT to be the optimal solution. The packing instance $B_{i}^{INS}$ corresponding to each OPT bin $B_i^{OPT}$ $\mbox{ for } i=1,\dots,q_t$ is as follows. 
\begin{enumerate}
\item The small items of each $B_{i}^{OPT}$ are copied to $B_{i}^{INS}$. Therefore, $B_{i}^{OPT}$ and $B_{i}^{INS}$ have same small items.

\item If $B_{i}^{OPT}$ has $p_j$ items from the set $L_{r,s}$ of the original instance $\mbox{ for } s=1, 2, \dots, \frac{1}{\epsilon^3}-1$ then $B_{i}^{INS}$ receives $p_j$ items from $L_{r,s+1}$ of the rounded up instance. We know that for each set $L_{r,s}$ in $L_{r}$, the cardinality of $L_{r,s}$ is either $\lfloor \epsilon^3\lvert L_r \rvert \rfloor$ or $\lceil \epsilon^3\lvert L_r \rvert \rceil$ $\mbox{ for } s=1,\dots, \frac{1}{\epsilon^3}$. Therefore, if we sometimes run out of items in some set $L_{r,s+1}$ and do not have an item to assign, it means that $\lvert L_{r,s}\rvert = \lvert L_{r,s+1}\rvert+1$. Such a value of $s$ exists only once per $L_i$ and is unique to that particular $L_i$. The space in the bin remains unused in this instance.
\end{enumerate}
\end{tcolorbox}
\begin{lemma}\label{pack}
If $OPT^{'}$ is the packing instance of $OPT$ then $OPT^{'} \leq OPT$.
\end{lemma}
\begin{proof}
The small items of $B_{i}^{INS}$ and $B_{i}^{OPT}$ remain unchanged. However, in the case where $B_{i}^{OPT}$ contains $p_j$ items from the set $L_{r,s}$ $\mbox{ for } s=1, 2, \dots, \frac{1}{\epsilon^3}-1$, then $B_{i}^{INS}$ will acquire $p_j$ items from $L_{r,s+1}$ of the rounded-up instance. Additionally, it is established that every item in $L_{r,s+1}$ has a lower weight than every item in $L_{r,s}$. Hence, the lemma is proved.
\end{proof}

\begin{figure}[t]
\includegraphics[width=14cm]{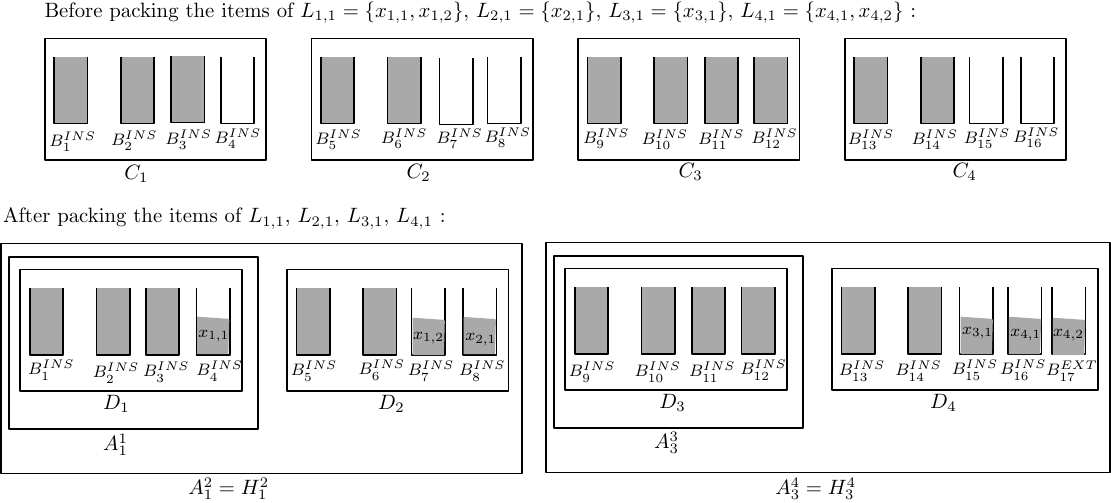}
\centering
\caption{Example of a house, apartment, room, and cell when $k=4$}\label{dp1}
\end{figure}
\begin{tcolorbox}[breakable,bicolor,
  colback=cyan!5,colframe=cyan!5,boxrule=0pt,frame hidden]

\noindent \textbf{\underline{Packing of Original Instance:}}
\vspace{4pt}
Although the packing instance associated with OPT has been established, it is evident that certain items have not yet been packed. When a bin $B_i^{OPT}$ contains $p_j$ items from the set $L_{r,s}$, then the contents of $B_{i}^{INS}$ are updated to include $p_j$ items from the rounded-up instance of $L_{r,s+1}$. This indicates that $\mbox{ for } r=1,\dots,k$, only the items in $L_{r,1}$ are not packed; therefore, $L=\cup_{r=1}^{k}L_{r,1}$ should be packed in the next formulation. Additionally, it is assumed that $B_{q_t}^{OPT}$ represents the final bin of OPT, then $B_{q_t}^{INS}$ is not necessarily the final bin in packing instance, as $B_{q_t}^{INS}$ can be empty. Therefore, it is assumed that the final bin of the packing instance is denoted by $B_{q_t^{'}}^{INS}$. Next, we define the \emph{cell}, \emph{room}, \emph{apartment}, and \emph{house}. Let $C_i=\{B_{\tau_i}^{INS}, B_{\tau_i+1}^{INS}, \dots, B_{\tau_{i+1}-1}^{INS}\}\mbox{ for } i=1,\dots,k-1$ and $C_k=\{B_{\tau_k}^{INS}, B_{\tau_k+1}^{INS}, \dots, B_{q_t^{'}}^{INS}\}$ be the cells. Each $C_i$ is copied into a new set $D_i$, so $D_i=C_i$, and this is known as a \emph{room} $\mbox{ for }i=1,\dots,k$ (see Figure \ref{dp1}). Cells are created before packing the items of $L$, but we must pack the items of $L$ in the rooms as follows. We first pack the items of $L_{1,1}$ left to right of the empty bins of $D_1$,\dots,$D_k$ so that there is exactly one item from $L_{1,1}$ in each empty bin. Similarly, we pack the items of $L_{i,1}$ left to right of the empty bins of $D_i$,\dots,$D_k$, making sure that every empty bin has exactly one item $\mbox{ for }i=2, \dots, k$. In some cases, if we run out of empty bins for $L_{i,1}$, we must introduce new bins after $B_{q_t^{'}}^{INS}$ and each new bin contains exactly one item of $L_{i,1}$ as before. We repeat this process $k$ steps to pack all the items of $L$. The extra bins are assumed to be $B_{q_t^{'}+1}^{EXT}$, \dots, $B_{q_t^{''}}^{EXT}$, we then assign all of these additional bins to $D_k$, that is $D_k=\{B_{\tau_k}^{INS}, \dots, B_{q_t^{'}}^{INS}, B_{q_t^{'}+1}^{EXT}, \dots, B_{q_t^{''}}^{EXT}\}$. A \emph{house} is defined as $H_1^{i}=\cup_{x=1}^{i}D_x$ if $L_{i,1}$ is the first set of $L$ such that $L_{i,1}$ became empty while placing the items in the empty bins of the set $D_i$. Subsequently, $H_{i+1}^{j}=\cup_{x=i+1}^{j}D_x$ is likewise a home if $L_{j,1}$ ($i<j$) is a second set of $L$ such that $L_{j,1}$ became empty when placing the items in the empty bins of the set $D_j$. We keep going through the same process until we get all the houses. It is clear that the number of houses is at most $k$. For a house $H_{i}^{j}$, $A_i^{r}=\cup_{x=i}^{r}D_x$ is called an \emph{apartment}$\mbox{ for } r=i,\dots,j$. It is evident that $H_{i}^{j}$ and $A_{i}^{j}$, in addition to $D_i$ and $A_i^{i}$, are identical. Assumed to be the set of all the houses is $H$, and the total cost of the bins in $H$ is \emph{cost(H)}. Take note that every item in $\cup_{l=i}^{j}G_l$ is packed in the house $H_{i}^{j}$, so any two houses can exist independently of one another.
\end{tcolorbox}

\begin{lemma} {\label{lemma 24}}
$Cost(H)\leq (1+c\epsilon)OPT^{'}$ where $OPT^{'}$ is the solution of packing instance and $c=\sum_{i=1}^{k}\frac{1}{4^{i-1}}\binom{k}{i}(4+9\tau_k)^i$.
\end{lemma}
\begin{proof}
The additional cost incurred for the original instance in comparison to the packaging instance is computed as follows. Suppose $H_{i}^{r}$ is a house, and $A_{i}^{i}$, $A_{i}^{i+1}$, \dots, $A_{i}^{r}$ are the blocks of $H_{i}^{r}$. We denote $M_l=\sum_{l=i}^{j}m_l$ and $L_q^{'}=\cup_{j=i}^{q}L_{j,1}$ $\mbox{ for } q=i,\dots,r$.

\vspace{10pt}

\noindent \underline{Inequality for $Cost(A_{i}^{i})$:}
\begin{equation}
\begin{split}
\Delta_i^{i} & = cost(A_{i}^{i})-cost(C_i)\\ &\leq \sum_{j=t_{i}+1}^{t_{i}+\lvert L_i^{'} \rvert} (j) \hspace{1cm}  (\text{where }B_{t_{i}}\text{ is the last bin in the cell }C_i)\\ & = (t_{i}+1)+(t_{i}+2)+\dots (t_i+\lvert L_i^{'}\rvert)  \\ & = t_{i}\lvert L_{i}^{'}\rvert+\frac{\lvert L_{i}^{'}\rvert(\lvert L_{i}^{'}\rvert+1)}{2} \hspace{1cm} (\text{use } M_i=\sum_{j=i}^{i}m_j=m_i \text{ and } \lvert L_{i}^{'}\rvert = \lvert L_{i,1}\rvert \leq 2M_i\epsilon^3 )  \\ & \leq 2M_i\epsilon^3t_{i}+3M_{i}^2\epsilon^6 \hspace{1cm} (\text{use lemma \ref{bubailemma2} and $(\lvert L_{i}^{'} \rvert +1) \leq (2M_i\epsilon^3+M_i\epsilon^3)$}).
\end{split}
\end{equation}
Now, we are looking at the least number of large items in $C_i$. The total number of large items in $C_{i}$ is at least
\vspace*{-\baselineskip}
\begin{equation}
\begin{split}
M_i-\sum_{j=i}^{i} L_{j,1} & \geq M_i-2(\sum_{j=i}^{i}m_j\epsilon^3)\\ &= M_i-2M_i\epsilon^3 \hspace{1cm}  (\text{where }M_i=\sum_{j=i}^{i}m_j=m_i)\\ & \geq \frac{31M_i}{32} \hspace{1cm}  (\text{since $\epsilon \leq \frac{1}{4}$}).
\end{split}
\end{equation}
Once more, $C_i$ contains a sum of $t_i$ non-empty bins containing a minimum of $\frac{31M_i}{32}$ large items. Therefore, ($t_i-\tau_i+1$) $\geq \frac{31M_i\epsilon^2}{32}$. Additionally, $C_i$ utilizes $t_i-\tau_i+1$ bins, and each of $t_i-\tau_i+1$ bins contains at least one item. Consequently, $Cost(C_i) \geq \frac{(t_i-\tau_i+1)^2}{4}$. Now,
\begin{equation}\label{inequality6}
\begin{split}
\Delta_i^{i} & = cost(A_{i}^{i})-cost(C_i) 
\\ & \leq 2M_i\epsilon^3t_{i}+3M_{i}^2\epsilon^6 
\\ & = 2M_i\epsilon^3(t_{i}-\tau_i+1) + 2(\tau_i-1)M_i\epsilon^3+3M_{i}^2\epsilon^6 \hspace{0.5cm} (\text{since } (t_i-\tau_i+1)+\tau_i-1 = t_i)
\\ & \leq 2\epsilon(t_{i}-\tau_i+1) \frac{32}{31} (t_{i}-\tau_i+1) + 2\epsilon(\tau_i-1) \frac{32}{31} (t_{i}-\tau_i+1) +3\epsilon^2 \left(\frac{32}{31} (t_{i}-\tau_i+1)\right)^2
\\ & \hspace{2cm} \hspace{0.5cm} \left(\text{Using } (t_i-\tau_i+1) \geq \frac{31M_i\epsilon^2}{32} \text{ and assume that } (\tau_i-1)\geq 0\right)
\\ & =  \frac{64}{31}\epsilon(t_{i}-\tau_i+1)^2 + \frac{64}{31} \epsilon(\tau_i-1) (t_{i}-\tau_i+1) +\frac{3072}{961}\epsilon^2 (t_{i}-\tau_i+1)^2
\\ & \leq \frac{64}{31}\epsilon 4Cost(C_i) + \frac{64}{31} \epsilon(\tau_i-1) 4Cost(C_i) +\frac{3072}{961}\epsilon \frac{1}{4} 4Cost(C_i)
\\ & \hspace{2cm} \hspace{0.5cm}  \left(\text{since } Cost(C_i) \geq \frac{(t_i-\tau_1+1)^2}{4} \geq \frac{(t_i-\tau_1+1)}{4} \text{ and }\epsilon \leq \frac{1}{4}\right)
\\ & \leq (4+9\tau_i)\epsilon Cost(C_i).
\end{split}
\end{equation}
Once more, the situation occurs when $(\tau_i-1)=-1$, or when $\tau_i=0$, the above inequality \ref{inequality6} will be 
\begin{equation}\label{inequality7}
\begin{split}
\Delta_i^{i} & \leq 2M_i\epsilon^{3}(t_i+1)-2M_i\epsilon^3+3M_i^{2}\epsilon^{6}
\\& \leq 2M_i\epsilon^{3}(t_i+1)+M_i^{2}\epsilon^{6} \hspace{2 cm} (\text{since } M_i^{2}\epsilon^6\geq M_i\epsilon^3\geq 1)
\\ & \leq 2\epsilon Cost(C_i). \hspace{2 cm} (\text{using the procedure used in the inequality \ref{inequality6} } )
\end{split}
\end{equation}
Combining inequalities \ref{inequality6} and \ref{inequality7} we conclude that $cost(A_{i}^{i})< \{1+(4+9\tau_i)\epsilon\} cost(C_i)$.
Similarly, the inequality for $Cost(A_{i}^{i+1})$ is as follows.

\vspace{10pt}

\noindent \underline{Inequality for $Cost(A_{i}^{i+1})$:}
\begin{equation}
\begin{split}
\Delta_i^{i+1} & = cost(A_{i}^{i+1})-(Cost(A_i^{i})+ cost(C_{i+1}))
\\ & \leq 2M_{i+1}\epsilon^3t_{i}+3M_{i+1}^2\epsilon^6 \hspace{0.5cm} (\text{where }M_{i+1}=\lvert L_{i+1}^{'}\rvert =\sum_{j=i}^{i+1}m_j=m_i+m_{i+1}).
\end{split}
\end{equation}
In the same manner as before, we are attempting to count the minimum number of large items in $A_i^{i}\cup C_{i+1}$. The sum of all large items in cup $A_i^{i}\cup C_{i+1}$ is equal to or greater than 
\begin{equation}
\begin{split}
M_{i+1}-\sum_{j=i}^{i+1} L_{j,1} & \geq \frac{31M_{i+1}}{32} \hspace{1cm}.
\end{split}
\end{equation}
Again, $A_i^{i}\cup C_{i+1}$ has a total of $t_{i+1}$ non-empty bins, with at least $\frac{31M_{i+1}}{32}$ large items. Therefore, ($t_{i+1}-\tau_i+1$) $\geq \frac{31M_{i+1}\epsilon^2}{32}$. Also, $A_i^{i}\cup C_{i+1}$ uses $t_{i+1}-\tau_i+1$ bins, and each of them has at least one item. Cosequently, $Cost(A^{i}_{i}\cup C_{i+1}) \geq \frac{(t_{i+1}-\tau_i+1)^2}{4}$.
Consequently, by employing the analogous inequalities of $A_i^{i}$, we obtain
Therefore, if we use the similar inequalities of $A_i^{i}$, then we get
\begin{equation}
\begin{split}
Cost(A_i^{i+1}) & \leq \{1+(4+9\tau_i)\epsilon\}Cost(A_i^{i}\cup C_{i+1}) \hspace{1cm}  (\text{using similar inequalities of }\Delta_i^{i})
\\ & = \{1+(4+9\tau_i)\epsilon\}\{(Cost(A_i^{i})+ Cost(C_{i+1})\}
\\ & \leq \{1+(4+9\tau_i)\epsilon\}[\{1+(4+9\tau_i)\epsilon \}Cost(C_i)+ Cost(C_{i+1})] 
\\ & \hspace{8cm} (\text{using inequality of }\Delta_i^{i})
\\ & = \{1+(4+9\tau_i)\epsilon\}^2Cost(C_i)+ \{1+(4+9\tau_i)\epsilon \} Cost(C_{i+1}) 
\end{split}
\end{equation}
Similarly, we must get 
\begin{equation}
\begin{split}
Cost(A_i^{i+2}) & \leq \{1+(4+9\tau_i)\epsilon\}^3Cost(C_i)+ \{1+(4+9\tau_i)\epsilon \}^2 Cost(C_{i+1})
\\ & \hspace{4.45cm }+ \{1+(4+9\tau_i)\epsilon \} Cost(C_{i+2})
\end{split}
\end{equation}
$$\vdots $$ 
Hence,
\begin{equation}
\begin{split}
Cost(A_i^{r}) & = Cost(H_i^{r}) 
\\ & \leq \{1+(4+9\tau_i)\epsilon\}^{r-i+1} Cost(C_i)+ \{1+(4+9\tau_i)\epsilon \}^{r-i} Cost(C_{i+1})
\\ & + \hspace{1cm} \{1+(4+9\tau_i)\epsilon \}^{r-i-1} Cost(C_{i+2})+\dots+\{1+(4+9\tau_i)\epsilon \} Cost(C_{r})
\\ & \leq \{1+(4+9\tau_k)\epsilon\}^{r-i+1} \{Cost(C_i)+Cost(C_{i+1})+ \dots + Cost(C_r)\}
\\ & \hspace{1cm} (\text{since }\tau_i\leq \tau_k \text{ and } \{1+(4+9\tau_i)\epsilon \}^{r-i+1} \geq \{1+(4+9\tau_i)\epsilon \}^{r-i-l} \text{ for } l\geq 0)
\end{split}
\end{equation}
Suppose, there are a total of $h$ houses $H_{1}^{l_1}$, $H_{l_1+1}^{l_2}$, \dots, $H_{l_{h-1}}^{l_h}$, where $l_1<l_2<\dots<l_h=k$. Therefore, $Cost(H_{1}^{l_1})+ Cost(H_{l_1+1}^{l_2})+ \dots+ Cost(H_{l_{h-1}}^{l_h}) = Cost(H)$, where $H$ is the union of the sets $H_{1}^{l_1}$, $H_{l_1+1}^{l_2}$, \dots, $H_{l_{h-1}}^{l_h}$, then we obtain
\begin{equation}
\begin{split}
 Cost(H) & \leq \{1+(4+9\tau_k)\epsilon\}^{l_1-1+1} \{Cost(C_1)+Cost(C_{2})+ \dots + Cost(C_{l_{1}})\}
\\ & + \{1+(4+9\tau_k)\epsilon\}^{l_2-l_1-1+1} \{Cost(C_{l_1+1})+Cost(C_{l_{1}+2})+ \dots + Cost(C_{l_{2}})\}
\\ & \dots
\\ & \{1+(4+9\tau_k)\epsilon\}^{l_k-l_{k-1}-1+1} \{Cost(C_{l_{k-1}+1})+Cost(C_{l_{k-1}+2})+ \dots + Cost(C_{l_{k}})\}
\\ & \leq \{1+(4+9\tau_k)\epsilon\}^{k} \{Cost(C_1)+Cost(C_{2})+ \dots + Cost(C_k)\}
\\ & \leq \{1+(4+9\tau_k)\epsilon\}^{k} OPT^{'}
\\ & \leq \left\{1+ \sum_{i=1}^{k}\frac{1}{4^{i-1}}\binom{k}{i}(4+9\tau_k)^i\epsilon \right\} OPT^{'} \hspace{1cm} \left(\text{ since } \epsilon \leq \frac{1}{4}\right)
\\ & = (1+ c\epsilon) OPT^{'} \hspace{1cm} \left(\text{ where } c=\sum_{i=1}^{k}\frac{1}{4^{i-1}}\binom{k}{i}(4+9\tau_k)^i \text{ is a constant }\right)
\end{split}
\end{equation}
\end{proof}
\begin{definition}
A packing that is possibly unfeasible but becomes feasible by eliminating the largest small item from each bin containing at least one small item is termed a quasi-feasible packing.
\end{definition}
\begin{definition}
Assuming a bin $B_j$ is packed at time $j$, its cost can be calculated as $j\times \lvert B_j \rvert$. This indicates that the cost per item of $B_j$ is denoted as $j$; we refer to this as the charging cost of an item of $B_j$. Therefore, the charging cost associated with each item in bin $B_j$ is denoted by $j$.
\end{definition}
A feasible packing is also a quasi-feasible packing, according to the definition. As a result, the cost of a quasi-feasible packing is either lower than or equal to the cost of a feasible packing. We now demonstrate that a cost factor of $(1+c_1\epsilon)$ must be applied in order to transform the quasi-feasible packing into a feasible packing, where $c_1=\frac{k+2}{4}+\frac{1}{2}$.
\begin{lemma}{\label{lemma25}}
    A quasi-feasible packing $Q$ can be converted into a feasible packing $P$ such that $Cost(P)\leq (1+c_1\epsilon)Cost(Q)$ where $c_1=\frac{k+2}{4}+\frac{1}{2}$
\end{lemma}
\begin{proof}
Suppose that the bins in $Q$ are $B_1$, $B_2$, \dots $B_{q_t^{''}}$, where first $\tau_1-1$ bins are empty. We have the rooms $D_i=\{B_{\tau_i}, B_{\tau_i+1}, \dots, B_{\tau_{i+1}-1}\}$  $\mbox{ for }i=1, \dots, k-1$ and when $i=k$ the room is $D_k=\{B_{\tau_k}, B_{\tau_k+1}, \dots, B_{q_t^{''}}\}$. We divide the room $D_i$ into multiple partitions of size $\frac{1}{\epsilon^2}$ bins from left to right, with the final partition possibly having fewer bins than $\frac{1}{\epsilon^2}$ bins. Let $D_i=\{P_i^{1}, P_{i}^{2}, P_{i}^{l_i}\}$ be the set of portions such that, with the exception of the final partition $P_{i}^{l_i}$, every $P_{i}^{j}$ has exactly $\frac{1}{\epsilon^2}$ bins. If $P_1^{1}$ contains fewer than $\frac{1}{\epsilon}-1$ bins, we add an extra bin after the final bin in $P_1^{1}$; otherwise, we add an extra bin at the $\frac{1}{\epsilon}$-th position in $P_1^{1}$. We add an extra bin at the beginning of each partition for every other partition (i.e., except $P_1^{1}$) of $D_i$ and $i\geq 1$. We find the largest small item in each bin for each partition and transfer it to the new bin for that partition. There are at most $\frac{1}{\epsilon^2}$ small items in each partition's corresponding new bin since each partition has at most $\frac{1}{\epsilon^2}$ bins, and each bin contains at most one largest small item. Each small item has a size of at most $\epsilon^2$, as we know. Thus, the new bin's total item size is limited to $\frac{1}{\epsilon^2}\times \epsilon^2=1$. As a result, each partition's new bin is a feasible bin. We reduce a quasi-feasible packing $Q$ to a feasible packing $P$ in this way. Additionally, it is evident that there are $k$ rooms in total, and each room has at least one partition. We now compute the additional cost resulting from the packing of each partition's largest small items into the new bin.

\smallskip

Initially, we compute the increment in charging cost for every item in $P_1^{1}$. Each bin's largest small item is moved to the new bin at $\frac{1}{\epsilon}$-th place. The charging cost of the largest item in the $i$-th bin is increased by $\frac{1}{\epsilon}-i$ when $i<\frac{1}{\epsilon}$. The charging cost of the largest item in the $i$-th bin does not increase when $i>\frac{1}{\epsilon}$. So, for the largest small items in each bin of $P_1^{1}$, the overall cost increment is therefore at most $\frac{\frac{1}{\epsilon}(\frac{1}{\epsilon}-1)}{2}\leq \frac{1}{2\epsilon^2}$. It is clear that the cost of $Q$ is at least $n$ for $n$ items. Using the observation \ref{bubai17}, we get $Cost(Q)\geq n \geq \frac{1}{\epsilon^3}$. Therefore, $\frac{\epsilon}{2}Cost(Q)$ is the maximum total cost increment for the largest small items in each bin of $P_1^{1}$. Once more, when $i<\frac{1}{\epsilon}$, the charging cost of all other items of the $i$-th bin in $P_1^{1}$ (i.e., excluding the largest small item of each bin) remains unchanged. However, when $i>\frac{1}{\epsilon}$, a new bin is placed, and the charging cost of all other items of the $i$-th bin in $P_1^{1}$ increases by exactly $1$. Moreover, the cost of every item in the $P_1^{1}$ in $Q$ is greater than or equal to $\frac{1}{\epsilon}$ for each $i$-th (where $i>\frac{1}{\epsilon}$) bin. As a result, the charging cost is increased by a multiplicative factor of at most $1+\epsilon$. The charging cost increment needs to be calculated for a partition $P_{i}^{j}$, where $i,j\geq 1$ except for $i=j=1$. In $Q$, the charging cost of every item in $P_i^{j}$ is at least $\frac{l_1+l_2+\dots +l_{i-1}+j-(i-1)}{\epsilon^2}=\frac{l-i+1}{\epsilon^2}$ because each item in the bins of $P_i^{j}$ has a charging cost increment of $l$, where $l=l_1+l_2+\dots+l_{i-1}+j$. Consequently, the charging cost is increased by a multiplicative factor of at most $1+\frac{l}{l-i+1}\epsilon^2\leq 1+(k+2)\epsilon^2\leq 1+\frac{k+2}{4}\epsilon$. Hence, combining all the cases, it is determined that $Cost(P)\leq (1+c_1\epsilon)Cost(Q)$, with $c_1=\frac{k+2}{4}+\frac{1}{2}$.
    
\end{proof}
\begin{definition}
Next Fit (NF) bin packing uses one active bin into which it packs the input; and once the free space in this bin becomes too small to accommodate the next item, a new active bin is opened and the previous active bin is never used again. Next Fit Increasing (NFI) bin packing is the same as NF, but the inputs are in increasing order with respect to their weights.
\end{definition}
\begin{definition}
 $NFI^{+}$ packs items sorted by non-decreasing
order as NFI, but it moves to the next bin only after the total weight of the contents of the current bin strictly exceed 1.
\end{definition}
\begin{definition}
For a feasible packing $P$, we define a quasi-feasible packing $Q_P$ as follows. Remove all small items in $P$ and re-pack them using $NFI^{+}$ into the empty spaces. Note that since $NFI^{+}$ is used here to pack only the small items, the resulting semi-feasible packing is indeed quasi-feasible packing as well.  
\end{definition}

\begin{lemma}\label{lemma27}
    $Cost(Q_P)\leq Cost (P)$
\end{lemma}
\begin{proof}
There is no repackaging involved in the large items. Therefore, the total cost of the large items in $P$ and $Q_P$ is the same. All we have to do is demonstrate that the entire cost of $Q_P$'s small items does not exceed the total cost of small items of $P$. Suppose that the number of bins used in $P$ is indicated by $y$. The number of small items in the $i$-th bin of $Q_P$ is represented by $q_i$, while the number of small items in the $i$-th bin of $P$ is represented by $p_i$. For $1\leq r\leq y$, we assert that $\sum_{i=1}^{r}q_i\geq \sum_{i=1}^{r}p_i$. Specifically, this would demonstrate that, given the number of small items being $\sum_{i=1}^{y}p_i$, the number of bins in $Q_P$ is at most $y$. Let $X_i$ represent the area in the $i$-th bin of $P$ that is not occupied by large items. In bins $1,\dots,r$ of $P$, the total size of small items is not greater than $\sum_{i=1}^{r}X_i$. Let $\zeta$ represent the final bin of $Q_P$ containing small items. 

\smallskip

Since the claim is evidently true for $r=\zeta, \dots, y$ if $\zeta \leq y$, we show it for bins $1,\dots, \min\{y,\zeta-1\}$. Since the following bin received at least one small item for each such bin of index $i$, the total size of all the small items in that bin in $Q_P$ exceeds $X_i$. Therefore, the total size of small items in bins $1,\dots,r$ of $Q_P$ exceeds $\sum_{i=1}^{r}X_i$ for every $r\in \{1,\dots, \min\{y,\zeta-1\}\}$. The claim follows since these bins in $Q_P$ contain the smallest little objects, while these bins in $P$ contain arbitrary items. Summing up all inequalities $\sum_{i=1}^{r}q_i\geq \sum_{i=1}^{r}p_i \mbox{ for }1\leq r\leq y$, we get $\sum_{r=1}^{y}\sum_{i=1}^{r}q_i\geq \sum_{r=1}^{y}\sum_{i=1}^{r}p_i$ implying that $\sum_{i=1}^{y}(y-i+1)q_i\geq \sum_{i=1}^{y}(y-i+1)p_i$ (by calculating the multiplier of every $q_i$ and every $p_i$ in the preceding sum). 
Thus, in $P$, the packing cost of the small items is $\sum_{i=1}^{y}i\times p_i$, whereas in $Q_P$, it is $\sum_{i=1}^{y}i\times q_i$. For any non-negative sequence $a_i$ ($1\leq i \leq y$), $\sum_{i=1}^{y}i\times a_i=\sum_{i=1}^{y}(t+1)\times a_i-\sum_{i=1}^{y}(t-i+1)\times a_i$ and since $\sum_{i=1}^{y} p_i= \sum_{i=1}^{y}q_i$, we get $\sum_{i=1}^{y}i\times q_i \leq \sum_{i=1}^{y}i\times p_i$ as required.
\end{proof}
\begin{lemma}
For any input I, the cost for the output of $NFI^{+}$ is not larger than $OPT(I)$.
\end{lemma}
\begin{proof}
The restrictions on weights of the small items are not used in Lemma \ref{lemma27}'s proof. The output of $NFI^{+}$ is $Q_P$ if all items are small (i.e., if $\epsilon=1$ is used), and $P$ is an optimal packing.
\end{proof}

We conclude from the last four lemmas that finding a quasi-feasible packing for the rounded-up instance is sufficient. Once such a packing is achieved, it can be further converted into a feasible packing for the rounded-up instance using Lemma \ref{lemma25}, and finally into a packing for the original instance using Lemma \ref{lemma 24}. Furthermore, according to Lemma \ref{lemma27}, each feasible packing has an equivalent quasi-feasible packing in which $NFI^{+}$ is used to pack small items into the spaces left by large items. With a constant factor $d$, which is independent of the input size, all these conversions can be carried out while keeping a cost that varies from the prior cost by a factor of $(1 + d \epsilon)$. For any packing $P$, we determine the quasi-feasible packing of the type $Q_P$ (this is achieved by dynamic programming in a directed graph, see below). This quasi-feasible packing is converted into a feasible packing, which is converted into a packing of the original items. 

\subsection{Dynamic Programming}

\begin{figure}[t]
\includegraphics[width=14cm]{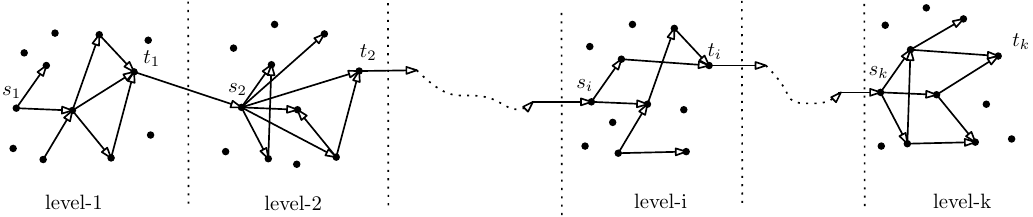}
\centering
\caption{An example of the graph $G(V,E)$ in dynamic programming is shown.}\label{dp2}
\end{figure}

Our dynamic programming algorithm is typically explained in terms of locating a path within a built-directed graph. In particular, we construct a directed graph $G = (V, E)$ in the manner shown below. Consider the labels as follows.
$$label(v_1)=(n_{1,2}(v_1), n_{1,3}(v_1), \dots, n_{1,\frac{1}{\epsilon^3}}(v_1), l_1(v_1))$$
$$label(v_2)=(n_{1,2}(v_2),\dots, n_{1,\frac{1}{\epsilon^3}}(v_2),n_{2,2}(v_2),\dots, n_{2,\frac{1}{\epsilon^3}}(v_2), l_2(v_2))$$
$$\vdots$$
$$label(v_k)=(n_{1,2}(v_k),\dots, n_{1,\frac{1}{\epsilon^3}}(v_k),\dots, n_{k,2}(v_k),\dots, n_{k,\frac{1}{\epsilon^3}}(v_k), l_k(v_k))$$
where $l_i(v_i)$ indicates that the small items $1$ through $l_i(v_i)$ are already packed and the items $l_i(v)+1$ through $\sum_{l=1}^{i}n_l-\sum_{l=1}^{i}m_l$ remain to be packed. $n_{i,j}(v_x)$ indicates the number of items of $L_{i,j}$ that are not packed yet (for $j\ge 2$). The $label(v_i)$ vector has a length of $i(\frac{1}{\epsilon^3}-1)+1$. For every potential label, we give a level. We designate $i$ as the level of $label(v_i)$. Consequently, the total number of levels is $k$. A directed graph $G(V, E)$ will have vertices in $V$ for each possible value of the label, where the directed edges must be formed next (see Figure \ref{dp2}).
\begin{lemma}
    $\lvert V\rvert$ is polynomial.
\end{lemma}
\begin{proof}
There are at most $m_i+1 \leq n_i+1$ different values that $n_{i,j}(v_x)$ may have. This is because the value of $n_{i,j}(v_x)$ is an integer in the interval $[0,m_i]$ for each value of $2\leq j \leq \frac{1}{\epsilon^3}$. The maximum number of distinct values for $l_i(v_i)$ is $\sum_{l=1}^{i}n_l-\sum_{l=1}^{i}m_l+1\leq \sum_{l=1}^{i}n_l+1$. As a result, the number of possible labels in $label(v_i)$ is at most $(\sum_{l=1}^{i}n_l+1)^{i(\frac{1}{\epsilon^3}-1)+1}\leq (n+1)^{i(\frac{1}{\epsilon^3}-1)+1}\leq (n+1)^{k(\frac{1}{\epsilon^3}-1)+1}$ and this is polynomial (as $\epsilon$ and $k$ are fixed constants).
\end{proof}
We next describe the edge set $E$. A vector for a generic packing pattern is defined as follows.
$$pat_1=(pat_{1,2}, \dots, pat_{1,\frac{1}{\epsilon^3}},p_1,l_1)$$
$$pat_2=(pat_{1,2}, \dots, pat_{1,\frac{1}{\epsilon^3}}, pat_{2,2}, \dots, pat_{2,\frac{1}{\epsilon^3}}, p_2,l_2)$$
$$\vdots$$
$$pat_k=(pat_{1,2}, \dots, pat_{1,\frac{1}{\epsilon^3}}, pat_{2,2}, \dots, pat_{2,\frac{1}{\epsilon^3}},\dots, pat_{k,2}, \dots, pat_{k,\frac{1}{\epsilon^3}},  p_k,l_k)$$
where $[p_i,l_i]$ represents the interval of the small items that are packed into this bin (that is, $p_i$, $p_{i}+1$, \dots, and $l_i$ are packed into this bin) and $pat_{i,j}$ indicates the number of items of $L_{i,j}$ that are packed into this bin. We require that for each $j$, $0\leq pat_{i,j}\leq \lvert L_{i,j}\rvert$, $1\leq p_i\leq l_i \leq \sum_{l=1}^{i}n_l-\sum_{l=1}^{i}m_l+1$, where $l_i=\sum_{l=1}^{i}n_l-\sum_{l=1}^{i}m_l+1$ only if $p_i=\sum_{l=1}^{i}n_l-\sum_{l=1}^{i}m_l+1$, in which case no small items are packed into the bin. Now, we introduce edges as follows. 

\smallskip

\begin{tcolorbox}[breakable,bicolor,
  colback=cyan!5,colframe=cyan!5,boxrule=0pt,frame hidden]

\noindent\textbf{\underline{Edges in level $i \mbox{ ($1\leq i\leq k$)}$ vertices:}}
\vspace{4pt}
Under the convention that $\sum_{l=p_i}^{p_i-1}s_l=0$, we need $\sum_{l=1}^{i}\sum_{j=2}^{\frac{1}{\epsilon^3}}(pat_{l,j}\times \sigma_{l,j})+\sum_{l=p_i}^{l_i-1}s_l\leq 1$ (keep in mind that $\sigma_{i,j}$ is the rounded up weight of each item of $L_{i,j}$ when $j\geq 2$ explained in \ref{roundedupsize}, and $s_l$ is the weight of the small item $p_l$), and if $l_i<\sum_{l=1}^{i}n_l-\sum_{l=1}^{i}m_l$, then $\sum_{l=1}^{i}\sum_{j=2}^{\frac{1}{\epsilon^3}} $ $(pat_{l,j}$ $\times \sigma_{l,j})+ \sum_{l=p_i}^{l_i}s_l> 1$. The first sum in these inequalities applies to sets of the form $L_{i,j}$, while the second sum applies to small items. The second condition is that $l_i<\sum_{l=1}^{i}n_l-\sum_{l=1}^{i}m_l$ (rather than $l_i\leq \sum_{l=1}^{i}n_l-\sum_{l=1}^{i}m_l$, since the bin containing the largest packed small item need not have a total weight greater than $1$). Consequently, the last bin containing small items will have the packing pattern with $l_i=\sum_{l=1}^{i}n_l-\sum_{l=1}^{i}m_l$, and other bins without any small items might exist, in which case $p_i=l_i=\sum_{l=1}^{i}n_l-\sum_{l=1}^{i}m_l+1$ would apply. Specifically, all packing patterns will have $p_i=l_i=1$ if no small items exist. A vector is not a generic packing pattern if it does not meet the above requirements. For every $j\geq 2$, if $pat_{r,j}\leq n_{r,j}(v_i) \mbox{ for }r=1,\dots,i \mbox{ and } j=2,\dots,\frac{1}{\epsilon^3}$ and $p_i=l_i(v_i)+1$, then a generic packing pattern $pat_i$ is said to be consistent with a vertex $v_i$. Additionally, for any generic packing pattern consistent with $v_i$, we have a directed edge running from $v_i$ to a vertex whose label is $(n_{1,2}(v_i)-pat_{1,2},\dots ,n_{1,\frac{1}{\epsilon^3}}(v_i)-pat_{1,\frac{1}{\epsilon^3}},n_{2,2}(v_i)-pat_{2,2},\dots ,n_{2,\frac{1}{\epsilon^3}}(v_i)-pat_{2,\frac{1}{\epsilon^3}},\dots, n_{i,2}(v_i)-pat_{i,2},\dots ,n_{i,\frac{1}{\epsilon^3}}(v_i)-pat_{i,\frac{1}{\epsilon^3}}, l_i )$ with the cost of the edge $\sum_{l=1}^{i}n_l-\sum_{l=1}^{i}m_l-l_i(v_i)+\sum_{l=1}^{i}\sum_{j=2}^{\frac{1}{\epsilon^3}}n_{l,j}(v_i)$ (that is, the number of items for which the ready time is less than or equal to $\tau_i$, that remain unpacked in $v_i$). 

\smallskip

We initially take a path from the vertex $s_1=(\lvert L_{1,2}\rvert, \dots, \lvert L_{1,\frac{1}{\epsilon^3}}\rvert, 0)$ and end it at $(0,\dots,0, n_1-m_1)$ if the path length is less than or equal to $\tau_2-\tau_1$. We then create some artificial vertices and directed edges (with the cost of the artificial edges are $0$) at the end of $(0,\dots,0, n_1-m_1)$ to get the length $\tau_2-\tau_1$. If the length of the path is greater than $\tau_2-\tau_1$, we start a path from the vertex $(\lvert L_{1,2}\rvert, \dots, \lvert L_{1,\frac{1}{\epsilon^3}}\rvert, 0)$ and end it at a vertex $t_1$ so that the length is $\tau_2-\tau_1$. Since $\lvert V\rvert$ is polynomial, finding $t_1$ must take polynomial time. We assume that $t_1=(\lvert L_{1,2}\rvert_{1}, \dots, \lvert L_{1,\frac{1}{\epsilon^3}}\rvert_{1},n_1^{*}-m_1^{*})$ where $\lvert L_{1,j}\rvert_{1}\leq \lvert L_{1,j}\rvert$ and $n_1^{*}-m_1^{*}\leq n_1-m_1$. We present $\tau_2-\tau_1$ bins, which correspond to the path of length $\tau_2-\tau_1$ mentioned above. Using the items in the aforementioned $\tau_2-\tau_1$ bins, we also update all packing patterns $pat_l$ $\mbox{ for }1=2, \dots, k$. From $t_1$ to $s_2=(\lvert L_{1,2}\rvert_{1}, \dots, \lvert L_{1,\frac{1}{\epsilon^3}}\rvert_{1},\lvert L_{2,2}\rvert, \dots, \lvert L_{2,\frac{1}{\epsilon^3}}\rvert,n_1^{*}-m_1^{*})$, we create a directed edge with cost $0$. $s_2$ is a vertex of level $2$.

\smallskip

For $2\leq i\leq k-1$, we take a path from the vertex $s_i$ and end it at $(0,\dots,0, \sum_{l=1}^{i}n_l-\sum_{l=1}^{i}m_l)$ (here, $i(\frac{1}{\epsilon^3}-1)+1$ is the vector length) if the path length is less than or equal to $\tau_{i+1}-\tau_{i}$ and then we create some artificial vertices and directed edges (with the cost of the artificial edges are $0$) at the end of $(0,\dots,0, \sum_{l=1}^{i}n_l-\sum_{l=1}^{i}m_l)$ to get the length of the path to be $\tau_{i+1}-\tau_i$; otherwise, we start at the vertex $s_i$ and follow a path to a vertex $t_i$, with the length $\tau_{i+1}-\tau_{i}$. We assume that $t_i=(\lvert L_{1,2}\rvert_{i}, \dots, $ $\lvert L_{1,\frac{1}{\epsilon^3}}\rvert_{i},\lvert L_{2,2}\rvert_{i}, \dots, \lvert L_{2,\frac{1}{\epsilon^3}}\rvert_{i},\dots,\lvert L_{i,2}\rvert_{i}, \dots, $ $\lvert L_{i,\frac{1}{\epsilon^3}}\rvert_{i},$ $ n_i^{*}-m_i^{*})$ where $\lvert L_{r,j}\rvert_{i}\leq \lvert L_{r,j}\rvert \mbox{ for } r=1,\dots,i \mbox{ and } j=2,\dots, \frac{1}{\epsilon^3}$. We present $\tau_{i+1}-\tau_i$ bins, which correspond to the path of length $\tau_2-\tau_1$ mentioned above. Using the items in the aforementioned $\tau_{i+1}-\tau_i$ bins, we also update all packing patterns $pat_l$ $\mbox{ for }1=i+1, \dots, k$. From $t_i$ to $s_{i+1}=(\lvert L_{1,2}\rvert_{i}, \dots, \lvert L_{1,\frac{1}{\epsilon^3}}\rvert_{i},\lvert L_{2,2}\rvert_{i}, \dots, \lvert L_{2,\frac{1}{\epsilon^3}}\rvert_{i},\dots,\lvert L_{i,2}\rvert_{i}, \dots, $ $ \lvert L_{i,\frac{1}{\epsilon^3}}\rvert_{i}, $ $\lvert L_{i+1,2}\rvert,\dots,$ $\lvert L_{i+1,\frac{1}{\epsilon^3}}\rvert, $ $ n_i^{*}-m_i^{*})$, we create a directed edge with cost $0$. $s_{i+1}$ is a vertex of level $i+1$.

\smallskip

For $i=k$, we take a path from the vertex $s_{k}$ and end it at $t_k=(0,\dots,0, \sum_{l=1}^{k}n_l-\sum_{l=1}^{k}m_l)$ (here, $k(\frac{1}{\epsilon^3}-1)+1$ is the vector length). We add one bin to each edge of the path from $s_1$ to $t_k$, with the exception of the edges between $t_i$ and $s_{i+1}$. This is our directed graph $G(V, E)$ as a result.
\end{tcolorbox}

\smallskip

\begin{tcolorbox}[breakable,bicolor,
  colback=cyan!5,colframe=cyan!5,boxrule=0pt,frame hidden]

\noindent\textbf{\underline{Explanation:}}
\vspace{4pt}
The path $\mathcal{P}$ from $s_1$ to $t_k$ provides us with a quasi-feasible packing on the original instance, which we will now explain. In the graph $G(V, E)$, let $\mathcal{P}_{i}$ denote the path from $s_i$ to $t_i$. A quasi-feasible packing of type $Q_{P_{i}}$ can be created given a path $\mathcal{P}_{i}$, and its cost is equal to that of the path $\mathcal{P}_{i}$. If the $l-$th edge of the path $\mathcal{P}_{i}$ corresponds to a pattern $pat_i$, then the $l-$th bin of the solution $SOL_i$ is created as follows. We pack the previously unpacked items from $L_{i,j}$ to this bin $pat_{i,j}$ for $j\geq 2$. If $p_i\leq \sum_{l=1}^{i}n_l-\sum_{l=1}^{i}m_l$, the set of small items $p_i$, $p_{i}+1$, \dots, $l_i$ are added to this bin. This solution $SOL_i$ is quasi-feasible since, by contradiction, the removal of the last (and largest) small item of each bin will leave items of total weight at most $1$. Note also that the cost of $SOL_i$ is exactly the cost of the path $\mathcal{P}_{i}$, since each item gets counted once for every bin up to the one that contains it. Conversely, in the graph with the same cost, a path from $s_i$ to $t_i$ in the quasi-feasible packing $Q_{P_{i}}$ corresponds to a feasible packing $P_i$. All bins are packed according to the above-defined pattern since the small items are packed using $NFI^{+}$. To be more precise, let's say we have a bin and an item set $S_i$ (remember that $S_i$ is the union of $G_i$ and the unpacked items of $G_1$,\dots, $G_{i-1}$, and initially $S_1=G_1$). $pat_{1,j}=\lvert L_{1,j} \cap S_i\rvert$ defines the first $\frac{1}{\epsilon^3}-1$ components of the pattern $pat_i$. Similarly, $pat_{2,j}=\lvert L_{2,j} \cap S_i\rvert$ defines the next $\frac{1}{\epsilon^3}-1$ components of the pattern $pat_i$. Also, $pat_{i,j}=\lvert L_{i,j} \cap S_i\rvert$ represents the $\left((i-1)(\frac{1}{\epsilon^3}-1)+1\right)$-th component to $i(\frac{1}{\epsilon^3}-1)$-th component of the same pattern $pat_i$. $p_i$ is the smallest index of any small item packed into the bin and $l_i$ is the largest index of any small item packed into the bin. If no small item is packed, then $p_i=l_i=\sum_{l=1}^{i}n_l-\sum_{l=1}^{i}m_l+1$. Once again, since each item gets counted once for every bin up to the one that contains it, the cost of the path is equal to the cost of the quasi-feasible packing $Q_{P_i}$. Since $\lvert V\rvert$ is polynomial, all these must be counted in polynomial time.

\smallskip

Observe that $s_1$ and $t_k$ are fixed vertices, as $s_1$ relates to every item in $G_1$, while $t_k$ relates to packing every item in $I$. Since there is an edge from $t_1$ to $s_2$, the path length from $s_1$ to $s_2$ is $\tau_2-\tau_1+1$. Thus, $\tau_k-\tau_1+(k-1)$ is the path length from $s_1$ to $s_k$. Since $\lvert V\rvert$ is polynomial and $\tau_k-\tau_1+(k-1)\leq \tau_k-k-2$ (a constant), the number of paths of length $\tau_k-\tau_1+(k-1)$ must be calculated in polynomial time. In order for the path from $s_k$ to $t_k$ to be the shortest path, we must find the optimal path from $s_1$ to $s_k$. After combining all $k$ steps, the shortest path in terms of cost corresponds to a quasi-feasible packing $Q_P$ of least cost, which we designate by $SOL$. This is because any quasi-feasible packing of the type $Q_P$ corresponds to a path $\mathcal{P}$. Recall that an output solution is constructed by converting $SOL$ into a feasible solution $SOL^{'}$ for the rounded-up instance, as shown in the lemma \ref{lemma25}, and converting $SOL^{'}$ into a solution $SOL^{''}$ to the original instance, as shown in lemma \ref{lemma 24}.
\end{tcolorbox}
\begin{lemma}
If $OPT$ is the optimal solution to our problem, the cost of $SOL^{''}$ is at most $(1+d\cdot\epsilon)OPT$ where $c=\sum_{i=1}^{k}\frac{1}{4^{i-1}}\binom{k}{i}(4+9\tau_k)^i$, $c_1=\frac{k+2}{4}+\frac{1}{2}$, and $d=c+c_1+\frac{c\times c_1}{4}$.
\end{lemma}
\begin{proof}
Let $OPT^{'}$ be the quasi-feasible packing of the rounded-up instance. Also, as previously noted, $SOL^{'}$ is the feasible packing of the rounded-up instance. Thus, $SOL^{'}$ provides a feasible solution for quasi-feasible packing $OPT^{'}$ for the rounded-up instance. Using Lemma \ref{lemma25}, we get $Cost(SOL^{'})\leq (1+c_1\cdot \epsilon) OPT^{'}$ where $c_1=\frac{k+2}{4}+\frac{1}{2}$. Again, using Lemma \ref{pack}, we obtain $Cost(SOL^{'})\leq (1+c_1\cdot \epsilon) OPT$. Also, using Lemma \ref{lemma 24}, we obtain $Cost(SOL^{''})\leq (1+c\cdot \epsilon)Cost(SOL^{'})$, where $c=\sum_{i=1}^{k}\frac{1}{4^{i-1}}\binom{k}{i}(4+9\tau_k)^i$. Furthermore, Lemma \ref{lemma27} states that each feasible packing has an equivalent quasi-feasible packing, where $NFI^{+}$ is utilized to pack small items into the spaces left by large items. Hence, $Cost(SOL^{''})\leq (1+c\cdot \epsilon)\cdot (1+c_1\cdot \epsilon)OPT = (1+(c+c_1+c\cdot c_1\cdot \epsilon)\epsilon)OPT \leq (1+d\cdot \epsilon) OPT$ where $d = c+c_1+\frac{c\times c_1}{4}$ (recall that $\epsilon \leq \frac{1}{4}$).
\end{proof}


\section{Concluding Remarks}

The evacuation problem is extended in this study with the help of the minsum bin packing problem, which takes items with different ready times and weights. We show APX-hardness for this new problem and apply it to develop an evacuation method for non-separable groups of individuals. Currently, our PTAS algorithm is restricted to a single destination dynamic edge. However, with minor modifications, this process works for a single destination path of any length. As an immediate direction for future work, the graph may be generalized to a tree while having a single destination. In this case, varying the edge capacities may drastically affect the evacuation process. It will be a much challenging task to obtain algorithmic results for graphs with multiple destinations.
%
%
\bibliographystyle{unsrt} 
\bibliography{references}  
\end{document}